\documentclass[12pt]{article}
\usepackage{graphicx}
\usepackage{epsfig}
\usepackage{amsmath}
\usepackage{amssymb}
\usepackage{amsthm}
\usepackage{graphics}
\def\beq{\begin{equation}}
\def\beqn{\begin{eqnarray}}
\def\eeq{\end{equation}}
\def\eeqn{\end{eqnarray}}
\def\lp{\left}
\def\rp{\right}

\def\eff{{\rm eff}}
\def\rc{r_c}

\def\ds{\displaystyle}

\def\dw{{dw}}
\def\lsim{\mathrel{\rlap{\lower3pt\hbox{\hskip0pt$\sim$}}
     \raise1pt\hbox{$<$}}}         %less than or approx. symbol
\def\gsim{\mathrel{\rlap{\lower4pt\hbox{\hskip1pt$\sim$}}
     \raise1pt\hbox{$>$}}}         %greater than or approx. symbol

\numberwithin{equation}{section}

\input epsf%
\begin{document}

\begin{flushright}
{NYU-TH-06/11/9}
\end{flushright}
\vskip 0.9cm

\centerline{\Large \bf  Domain Walls As Probes Of Gravity}

\vspace{1cm}

\centerline{\large Gia Dvali\footnote{gd23@nyu.edu}, Gregory
Gabadadze\footnote{gg32@nyu.edu }, Oriol
Pujol{\`a}s\footnote{pujolas@ccpp.nyu.edu} and Rakibur
Rahman\footnote{mrr290@nyu.edu}}

\vspace{5mm}

\centerline{\emph{Center for Cosmology and Particle Physics}}
\centerline{\emph{Department of Physics, New York University} }
\centerline{\emph{New York, NY, 10003, USA} }

\vspace{1cm}

\begin{abstract}

\vspace{0.1in}

We show that domain walls are probes that enable one to
distinguish large-distance modified gravity from general
relativity (GR) at short distances.  For example, low-tension
domain walls are stealth in modified gravity, while they do
produce global gravitational effects in GR. We demonstrate this by
finding exact solutions for various domain walls in the DGP model.
A wall with tension lower than the fundamental Planck scale does
not inflate and has no gravitational effects on a 4D observer,
since its 4D tension is completely screened by gravity itself. We
argue that this feature remains valid in a generic class of models
of infrared modified gravity. As a byproduct, we obtain exact
solutions for super-massive codimension-2 branes.

\end{abstract}

\newpage

\section{Introduction and Summary}

A model of large distance modification of gravity \cite{dgp}
is based on the existence of a new mega-scale in the theory,
$r_c\sim H_0^{-1}\sim 10^{28}~{\rm cm}$.  In this model, the graviton
propagates extra degrees of freedom (two helicity-1
and one helicity-0 states), that lead to the
van Dam-Veltman-Zakharov (vDVZ) \cite{vDVZ}
type discontinuity in the linearized approximation.

However, the continuity in recovering the Einsteinian metric for
localized sources in the $r_c \rightarrow \infty$  limit, is due
to non-linear interactions of these  extra states, that become
strongly coupled,  and  `self-shield' at short distances
\cite{ddgv}.

Because of  these non-linearities, any spherically-symmetric
source of a Schwarzs\-child radius $r_g \lsim r_c$, gets endowed
with a new physical distance scale $r_*\sim (r_g r_c^2)^{1/3}\lsim
r_c$. The modifications to the metric now depend on whether the
source is  bigger or smaller than $r_*$: For sources localized
well within their own $r_*$, the metric inside the $r_*$-sphere
($r \ll r_*$) is almost Einsteinian, with tiny corrections, which
can be computed either in the $1/r_c$-expansion
\cite{Gruzinov,dgz,takahiro} or exactly \cite{gi}.

For the existing realistic localized sources, such as stars, or
planets (or even galaxies) the values of $r_*$ are huge, and
correspondingly, the deviations from the Einsteinian metric are
strongly suppressed near the sources. One expects that the
corrections are potentially measurable  \cite
{dgz,luestarkman,iorio,gi} by Lunar laser ranging experiments
\cite{Adelberger}, but in general, detection of the modifications
at short scales is extremely hard. In this respect, it is
important to identify other types  of gravitating sources, for
which modifications are significant, even at the short distances.

We shall show that domain walls, are exactly these type of sources.
In particular, we will find that low tension walls do not gravitate
in the DGP model \cite {dgp}.  We will  also argue that the latter
property should persist in other theories of large distance modified
gravity.

Another important question concerns  the form of the metric beyond
$r_*$.  The naive perturbative expansion suggests that for $r_*
\ll r \ll r_c$ the metric should have an approximately
four-dimensional, $1/r$, scalar-tensor-gravity type form
\cite{dgp,ddgv}. However, a non-perturbative solution of Ref.
\cite{gi} exhibits a different behavior: for $r \gg r_*$ the
metric turns into the one produced by a five dimensional source.
This is because the four-dimensional mass of the source is
completely screened by a halo of non-zero curvature that surrounds
the source and extends to the distances $\sim r_*$  \cite {gi}.

It is important to understand the significance and universality of
the screening mechanism \cite {gi}. The question whether the
direct crossover, beyond $r_*$, into the five dimensional regime,
is a general property of well-localized sources, should be
understood better.  Although, in the present work we won't be able
to clarify this issue completely, interestingly enough, we will
find a similar screening mechanism  for the domain walls.  As a
result, properties of the domain walls are dramatically different
in modified gravity, and they could serve as very effective probes
to distinguish between modified gravity and general relativity (GR).

Consider 4D space-time labelled by coordinates
$(t,x,y,z)$ and introduce a Nambu-Goto domain wall  with the
stress-tensor
\beq
T_{\mu\nu}=\sigma \, \delta(z) \;{\rm diag}(1,-1,-1,0)~,
\label{dwt}
\eeq
where $\sigma$ is the tension of the wall and $z$ denotes the
coordinate transverse to its worldvolume.  For simplicity, we
consider the wall to be straight in this coordinate system
and localized at the point $z=0$. For the moment we
ignore  the transverse size of the wall. Gravitational field
of such a source was found by Vilenkin \cite {Vilenkin} and by
Ipser and Sikivie \cite {IS} (VIS),  and reads  as follows:
\beq
ds^2 = (1-H|z|)^2\left (-dt^2 + e^{2Ht}(d{x}^2 +dy^2)\right ) +dz^2\,,
\label{VIS}
\eeq%
where $H=2\pi G_N \sigma$. One simple consequence of this metric
is that a freely falling observer will be  repelled from the wall.
In general,  an observer outside the wall floats in the Rindler
space while on the wall he/she observes a 3D de Sitter expansion.

   Using the above knowledge, and trying to guess the behaviour of the
domain walls in DGP, one comes to the following puzzle. Naive
intuition would tell us that, at sufficiently short distances, the
wall on the brane should behave as in 4D, and thus, inflate. On
the other hand, at large distances, the  wall should behave as a
codimension-two object, which are known to produce a static metric
with just a deficit angle, at least for sufficiently small
tensions. How could a non-singular metric interpolate between the
inflating and static patches?

As we shall see, the resolution of this puzzle lies in the
dramatically different behaviour of walls in GR and in DGP-like theories.
In modified gravity such
a domain wall becomes {\it stealth}, as long as its tension does
not exceed a certain critical value. The wall will have no
gravitational effects whatsoever. The presence of such a wall can
only be detected by scattering on its core.  For instance, if
gravity is modified at cosmological distances, there could exist a
domain wall passing through the Solar system; we would be able to
discover  it only  by coming into a contact with its core. On the
other hand, a long-range field of such a domain wall would be felt
in the Solar system if gravity is described by GR.

Detailed explanations of why this takes place is given in the bulk
of the paper. Here we discuss one way to understand this in the
context of DGP.  Consider the modified Einstein equation:
\beq
G^{(4)}_{\mu\nu} + m_c (K_{\mu\nu} - g_{\mu\nu}K) = 8\pi G_N
T_{\mu\nu}\,, \label{ModEin}
\eeq
where $m_c\equiv r_c^{-1}$,
$G^{(4)}_{\mu\nu}$ is a 4D Einstein tensor, $K_{\mu\nu}$ is an extrinsic
curvature tensor of 4D space-time, and $K$ is its trace (for the
full set of equations see, e.g., \cite {gi}). The second term on
the l.h.s. of Eq. (\ref {ModEin}) is what distinguishes it from
the conventional Einstein equation. Moreover, (\ref {ModEin})  can
be interpreted as a GR equation with an effective source  $T^{\rm
eff}_{\mu\nu} = T_{\mu\nu} - {m_c \over 8\pi G_N } (K_{\mu\nu} -
g_{\mu\nu}K)$. For the domain wall the
extrinsic curvature terms exactly compensate the stress tensor
(\ref {dwt}) giving rise to $T^{\rm eff}_{\mu\nu} = 0$.  Hence,
the tension of the wall, as seen from the point of view of a 4D
observer, is screened entirely by gravitational effects encoded in
the extrinsic curvature. Not surprisingly, the domain wall worldvolume
remains flat, and so does the metric on the brane:
\beq%
ds^2|_{\rm 4D} = -dt^2 + d{x}^2+dy^2+dx^2\,.
\label{4D metric}
\eeq%

Furthermore, we shall see that this screening
takes place inside the core of the wall. Hence, the analog notion of the
$r_*$ scale for a domain wall (understood as where the self-shielding takes place)
coincides with its thickness,
$$
r_*^{(dw)}=d_{core}~.
$$
This is to be compared to the Schwarzchild-like case \cite{gi}, where
the shielding also occurs, and $r_*$ extends outside the source.
The net result is the screening of the 4D tension/mass in both cases.

Although the above arguments
were given for the DGP model, as we show in Section \ref{sec:gia},
a similar mechanism is expected to remain valid generically in
models of long distance modified gravity in which the graviton is
no longer massless \cite{dvali06} and acquires extra
polarizations.

\section{Domain Walls in Perturbation Theory}
\label{sec:lin}

In this subsection we analyze the domain wall metric in
perturbation theory. Subtleties of such an approach do manifest
themselves already in the case of GR, as has been discussed in
\cite {Vilenkin0}. Here, we reiterate and generalize those arguments.
Consider again 4D space-time and a domain wall in it. Its metric
is given by (\ref {VIS}). Can this metric be recovered in a
perturbative calculation? This question was addressed in Ref.
\cite {Vilenkin0} and  we shall briefly review the argument for
clarity.

Given that the equation for the linearized metric is second order
in derivatives, and from the symmetries of the source (\ref{dwt}),
one {\it naively} obtains the result that the linearized metric
potentials behave like $\propto |z|$. In appropriate coordinates,
the metric takes the form
\beq %
ds^2|_{\rm Pert.~Th.}\simeq (1-2H|z|)\left (-dt^2 + d{x}^2
+dy^2\right ) +dz^2\,, \label{VISpt}
\eeq%
where $H=\sigma/4m_P^2$. This is consistent with the
linearizatioin of the exact solution (\ref{VIS}),
\beq%
ds^2|_{\rm Lin.}\simeq (1-2H|z|)\left (-dt^2 + (1+2Ht)(d{x}^2
+dy^2)\right ) +dz^2\,. \label{VISptOK}
\eeq%
Indeed, one can check that
(\ref{VISpt}) and (\ref{VISptOK}) differ by a gauge transformation
$\partial_{(\mu}\xi_{\nu)}$ with $\xi^\mu$ of the form $\xi^z=0$,
$\xi^0=H(x^2+y^2)/2 $ and $\xi^i=H\,t\,x^i$ (here $x^i$ label $x$
and $y$)\footnote{Note, though, that this transformation diverges
at infinity. Hence, (conserved) sources
that do not decay at infinity
can probe the difference between (\ref{VISpt}) and (\ref{VISptOK}).}.
Hence, to linear order in
perturbation theory, the metric does not capture the fact that the
wall inflates, which makes sense since the curvature scalar of the
worldvolume in the full solution is of order $H^2$.
The general lesson to be learned from the above considerations is
as follows:
it seems that the linearized theory does not capture all the physics, and
one needs to go  to higher orders in perturbation
theory (or possibly to the full nonlinear theory).\\

Now, let us perform the perturbative calculations
for a domain wall localized on the brane in the DGP model.
Here, we consider the walls with a sub-critical tension
$\sigma  \lsim M_*^3$. The subtleties outlined above
manifest themselves in this case too. The
linearized equations have a family of solutions,  some of them
are  static  and some are time dependent. It is hard to guess
{\it a priori} what the right linearized solution is. However,
we know from the exact solution of  the next section
the following:  Unlike the domain wall in 4D GR,
the worldvolume of a domain wall with $\sigma  \lsim M_*^3$ in
DGP does not inflate. Hence, we can choose accordingly the
right linearized solution\footnote{Note that we would obtain
a wrong linearized solution if we did not know the exact solution
and we were to follow  the intuition gained on domain walls
in 4D GR.}. Taking the above arguments  into account,
we now turn to the perturbative considerations.

Expanding the metric around the 5D Minkowski vacuum
$$
g^{(5)}_{MN}=\eta_{MN}+h_{MN}~,
$$
and using the harmonic gauge in the bulk, we obtain
that the only nonzero components, $h_{\mu\nu}$ and $h_{55}$,
satisfy the relation
$$
h\equiv h_\mu^\mu=h_5^5 ~,
$$
where the brane is located at $x_5=0$.
The linearized Einstein's equations take the form \cite{dgp}
\beq\label{harmonic}%
\lp({M_*^3\over2}\,\Box_5+{m_P^2\over2}\,\delta(x_5)\,\Box_4\rp)
\;h_{\mu\nu} =- \delta(x_5)\lp(T_{\mu\nu}-{1\over3}T
\eta_{\mu\nu}\rp)+{m_P^2\over2}\partial_\mu\partial_\nu h ~,%
\eeq%
while the trace of this equation is
\beq\label{trace} %
{M_*^3\over2}\Box_5 \; h=  {1\over3}\,\delta(x_5)\;T ~.%
\eeq %
For a domain wall $T=-3\sigma \delta(z)$. By the symmetries of the
problem, the {\it static} solution should depend only on $|x_5|$
and $|z|$, and satisfy $\partial_5 h=0$ for $x_5=0$ (and
$z\neq0$).
One can show that up to a constant, the only such a
solution with the appropriate singular behavior at $z=x_5=0$ is
\beq\label{h} %
h=- {1\over 2\pi} {\sigma\over M_*^3} \log\lp[(z^2+x_5^2)\mu^2\rp] ~, %
\eeq %
where $\mu$ is a mass scale, which when the transverse size $d$ is incorporated
turns out to be of order $1/d$.
Once we found $h$, we can plug it into the r.h.s. of
(\ref{harmonic}), to find the full $h_{\mu\nu}$. Now, since on the
brane $h$ depends only on $z$, one immediately sees that the
components parallel to the DW have no source term. Requiring that
they are regular at infinity, one obtains that they should vanish.
Thus, $h=h_z^z$, and indeed the $(zz)$ equation is trivially
satisfied. Hence, to leading order in $\sigma/M_*^3$, the metric
is
\beq\label{linsolution} %
ds^2=(1+h)(dx_5^2+dz^2)-dt^2+dx^2+dy^2 ~,
\eeq %
with $h$ given in (\ref{h}). The metric (\ref{linsolution}) can be brought to
a locally flat form
\beq\label{linsolutionProper} %
ds^2=d\rho^2+ \rho^2 d\theta^2 -dt^2+dx^2+dy^2 ~,
\eeq %
with the range of the angular coordinate modified (to leading order)
as $0\leq\theta<2\pi-\delta$ with
$$
\delta = {\sigma\over M_*^3}\,.
$$
This is the well known conical space induced
by a local codimension-2 object, with deficit angle given by $\delta$.
Interestingly, this feature holds true also in
the exact solutions for \emph{sub-critical} domain walls (those
with $\sigma < M_*^3$) on the conventional branch, as we shall see
in Sec \ref{sec:CB}. Let us remark now that (\ref{linsolution})
implies that the induced metric on the brane is flat; setting
$x_5=0$ and using the coordinate $dz_*\equiv\sqrt{1+h}\,dz$, the
induced metric is $dz_*^2-dt^2+dx^2+dy^2$. Hence, a (sub-critical)
domain wall in DGP has no gravitational effect on the brane. As
will become more clear below, the wall switches on the extrinsic
curvature term in such a way that the latter behaves as a negative
tension wall, screening it completely.

\section{Exact solutions}
\label{sec:exact}

The DGP action \cite {dgp}  of the bulk plus brane plus the Domain
Wall (DW) in the thin wall approximation takes the form:
\beq\label{action}%
S=\int d^5x \sqrt{-g_{(5)}} \;{M_*^3\over2} \; R_{5} %
%+\int d^4x \sqrt{-h} \lp(\;{m_P^2\over2} \,R_{4} -\tau \rp)%
+\int d^4x \sqrt{-g} \;\lp( {m_P^2\over2} \;R_{4} -\tau\rp) %
-\int d^3x \sqrt{-\gamma} \;\sigma \,, %
\eeq%
where $R_5$ and $R_4$ are the Ricci scalars of the metric in the
bulk $g^{(5)}_{\mu\nu}$ and of the induced metric on the brane
$g_{\mu\nu}$, and $\gamma_{\mu\nu}$ is the induced metric on the
DW. We have also introduced the brane tension $\tau$, even though
our main interest is in the case $\tau=0$.
In the derivation of the exact solutions below, we follow
\cite{gp}.
Given its potential relevance for the present acceleration of the universe
\cite{perl,riess}, we shall discuss both the conventional and the self-accelerated
branches of the theory \cite{cedric,ddg}.

The equations of motion arising from (\ref{action}) can be split
into the equations for the bulk and those for the brane. We assume
that there is no cosmological constant in the bulk, and that it
has at least the same symmetries as the (maximally symmetric)
domain wall. We are mostly interested in the cases  when (1) the
DW is inflating, in which case its geometry is a 3 dimensional de
Sitter space; (2) the DW is flat. Then, the analog of Birkhoff's
theorem in 5D ensures that the bulk metric locally has the form
\beq\label{Cmetric} %
ds^2=f(R)dZ^2 + {dR^2\over f(R)} +R^{2} ds_{u}^2 ~,
\eeq %
with $f(R)=u - C_u/R^2$ and $C_u$ an  integration constant. Here,
$ds_{u}^2$ is the line element of a three dimensional Minkowski
space ($u=0$) or de Sitter space with unit curvature radius
($u=1$). In this article, we shall consider only those boundary
conditions in the bulk that fix it to be flat, that is $u=1$,
$C=0$. The general case will be discussed elsewhere \cite{dgprRH}.
Still, it is convenient to introduce two different slicings,%
\beq%
\label{bulkmetric} %
ds^2=dZ^2 + dR^2 +R^{2\kappa} ds_{\kappa}^2,
\eeq%
corresponding to the full Minkowski  space ($\kappa=0$) and  to a
kind of Rindler space ($\kappa=1$) .
As we will see shortly, the 3-dimensional geometries of the domain
wall are given by $ds^2_\kappa$. Hence, the Rindler (Cartesian)
coordinates are well suited for an inflating (flat)  wall.

The equations for the brane are given by the Israel junction
conditions,
\beq%
\label{israel} %
2 M_*^3 K_{\mu\nu}=T^\eff_{\mu\nu}-{1\over3}T^\eff g_{\mu\nu}\,,
\eeq%
where we imposed $Z_2$ symmetry across the brane, $K_{\mu\nu}$ is
the extrinsic curvature, $T^\eff\equiv T^{\eff~\mu}_{~~~\mu}$  and %
\beq%
T^\eff_{\mu\nu}=-m_P^2 G^{(4)}_{\mu\nu} -\tau g_{\mu\nu} -\sigma
\delta(\xi) \gamma_{\mu\nu}\,.
\eeq%
Here,  $\xi$ is the coordinate along the brane orthogonal to the
DW, and $\tau$ is the brane tension.

The idea is now the following. Once the form of the bulk is known,
the problem reduces to finding the embedding of the brane in this
space, via Eq. (\ref{israel}).
The location of the brane can be parameterized by two functions
$(R(\xi),Z(\xi))$. Then, the induced metric on the brane is
\beq\label{induced}%
ds_4^2= %(f(R)Z'^2+{R'^2\over f(R)})
d\xi^2+R^{2\kappa}(\xi)\;ds_{\kappa}^2\,,
\eeq%
where we have chosen the `gauge' %
\beq\label{Z} %
Z'^2+R'^2=1~,
\eeq%
and $'$ denotes a $\xi$-derivative. This can be thought of
as fixing the arbitrariness introduced in the parameterization
$(R(\xi),Z(\xi))$. We then compute Eqs. (\ref{israel}) explicitly
in terms of this parameterization, which allows us to solve for
$R(\xi)$ (and $Z(\xi)$, through (\ref{Z})).
Generically, the brane will `cut' the $Z-R$ plane in two regions,
and the full 5D space  is then obtained by taking two identical
copies of the appropriate side pasted together along the brane.

The components
of (\ref{israel}) along the 3D de Sitter/Minkowski directions lead to
\beq%
\label{angular}%
\epsilon \;\kappa\; {\sqrt{1-R'^2}\over r_c \, R} =
\;-\kappa\;{1-R'^2\over R^2}+{\tau\over3m_P^2}~,
\eeq%
where
\beq\label{rc} %
r_c={m_P^2\over 2M_*^3} ~.%
\eeq%
The parameter $\epsilon=\pm1$ is the sign of the extrinsic
curvature. %, that is, the sign of the effective tension on the
%brane.
%
For $\tau>0$, $\epsilon$ is positive (negative) for the
Conventional (Self-Accelerated) branch.

Equation (\ref{angular}) leads to %
\beq\label{r'r}%
%{1-R'^2\over R^2}=H^2
{1-R'^2\over R^2}\equiv H^2~,
\eeq%
where $H$ solves
\beq%
\label{H} %
\epsilon \; \kappa\; {H\over r_c } = -\kappa \;
H^2+{\tau\over3m_P^2}~.
\eeq%
The case $\kappa=0$, can only be met for $\tau=0$, which makes
perfect sense, since in this case the metric (\ref{induced}) is
flat. For $\kappa=1$, the solutions symmetric across the DW are of
the form
\beq%
%{1-R'^2\over R^2}=H^2
\label{Rxi}%
R(\xi)=H^{-1}\sin\lp[H(\xi_0-|\xi|)\rp]
\eeq%
where we have introduced the integration constant $\xi_0$ so that
the DW is placed at $\xi=0$.\\

The  $(\xi\xi)$ component of (\ref{israel}), involving the
localized source term $\sim \sigma\delta(\xi)$ determines the
constant $\xi_0$.
This equation is not independent of Eq. (\ref{angular}), which is
valid away from the source  localized at $\xi=0$.
Close to the source we obtain the following divergent terms
\beq%
\label{xixi1}%
-\epsilon {R''\over 2r_c\sqrt{1-{R'}^2}} = \kappa \; {R''\over R}
+{2\over r_\dw} \delta(\xi) ~, %+{2\over3}{P_\xi\over 2m_P^2}~,
\eeq%
where
\beq\label{rg} %
r_\dw\equiv{4m_P^2\over \sigma} ~,%
\eeq%
is the would-be  domain wall horizon.
Integrating both sides of (\ref{xixi1}) from $\xi=-\varepsilon$ to
$\xi=\varepsilon$, and letting $\varepsilon\to0$, one obtains the
following matching condition for the jump of $R'$,\footnote{Note
that in Ref. \cite{gp} the integration of the l.h.s. of
(\ref{xixi}) is incorrectly done.}
\beq%
\label{xixi}%
-{\epsilon\over 2 \rc} \Delta \arcsin R'_0 =\kappa\;{\Delta
R'_0\over R_0} +{2\over r_\dw}\,,
\eeq%
where $\Delta X\equiv \lim_{\varepsilon\to0}
[X(\xi+\varepsilon)-X(\xi-\varepsilon)]$. To arrive at (\ref{xixi}),
we have used that $R$ is continuous at $\xi=0$, and
it can be treated as a constant for $\xi$ close enough to $0$.

\begin{figure}[!tb]
\begin{center}
  \includegraphics[width=8cm]{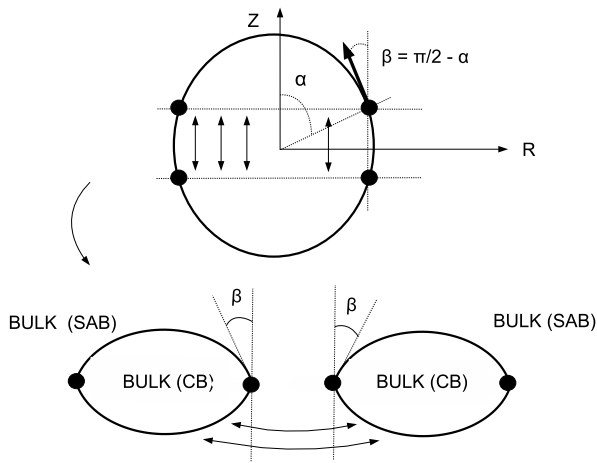}%
\caption{The brane trajectory in the $R-Z$ plane, for a generic DW
in an inflating brane $(H\neq0)$. The two figures represent a
constant time snapshot, and one of the directions along the DW
(represented by the thick dot) is obtained in the upper diagram by
rotating it around the $Z$ (vertical) axis. The space is cut by the
two constant $Z$ sections where the brane is at an angle
$\alpha=H\xi_0$. The space in between is then removed, and the two
slices joined together. Then, another identical copy of the
resulting `pancake' is glued along the brane, as indicated in the
lower diagram. For the conventional branch (CB), the bulk is the
interior, while for the self-accelerated branch (SAB), it is the
exterior. For the CB, this space has a deficit angle
$\delta=4\beta$, with $\beta=\pi/2-\alpha$. For the SAB, we have
an \emph{excess} angle,  that is $\delta=-4\beta$.}
\label{fig:Z2SAB} %
\end{center}
\end{figure}

Now, the gauge condition (\ref{Z}) can interpreted as the
condition that the vector $(R',Z')$ has unit norm. We can thus
identify $|R'_0|=\sin\beta$. From Fig. \ref{fig:Z2SAB}, $\beta$ is the
angle between the tangent to the brane and the $Z$ axis at the DW location. Upon
cutting and pasting the appropriate sections of the $R-Z$ plane,
we realize that $\beta $ is related to the deficit angle (see Fig
\ref{fig:Z2SAB}) through
\beq\label{deltabeta} %
\delta=4 \epsilon \beta~.
\eeq %
Note that the sign of $\delta$ tells us whether it is a deficit or
excess angle. Equation (\ref{xixi}) then takes a simple
form\footnote{This correctly reproduces the 4-dimensional and
5-dimensional limits. For $\rc\to\infty$, we obtain $ -{\Delta
R'_0/ R_0}={2/ r_\dw}~, $ which we identify as the Israel junction
condition for a DW in GR. For $\rc=0$, we get $
\delta=\lim_{\rc\to0}{8\rc/ r_\dw}={\sigma/ M_*^3}~, $ which is the
correct relation between the tension and the deficit angle.}
\beq%
\label{xixideficit}%
-\kappa\;{\Delta R'_0\over R_0}+\,{\delta\over 4 \rc}={2\over
r_\dw}\,.
\eeq%
The interpretation of this is clear. It tells us how much the DW
is behaving as a codimension-2 object (opening-up  a deficit angle),
{\it versus} a codimension-1 object embedded on the brane worldvolume
(generating a jump in the derivative of the induced metric).

Assuming $Z_2$ symmetry across the wall
Eq (\ref{xixideficit}) reads, in terms of $\beta$,
\beq\label{xixiBeta} %
2\kappa\,H\rc\,{\tan\beta} +\epsilon \; \beta=%{2\rc\over r_\dw} =
{\sigma\over 4 M_*^3} ~.
\eeq%
This equation cannot be solved analytically (except for
$\kappa=0$), but it is straightforward to solve it numerically for
given values of the parameters. We shall now do this in the next
subsections.

For clarity, we shall now summarize the main features of these
solutions. It is already apparent from (\ref{xixideficit}) that
both the 5D effect (amount of deficit angle) and the 4D effect
(the jump within the brane induced metric) produced by the wall
will in general differ from the usual pure 4D or 5D ones. In order
to quantify these deviations we can always introduce the effective
4D and 5D tensions in terms of whichever geometrical effect is
produced,
\beq\label{4Dsigma} %
\sigma^{(4D)}_\eff \equiv -2 \kappa \;m_P^2\;{ \Delta R'_0\over
R_0}
\eeq%
and
\beq\label{5Dsigma} %
\sigma^{(5D)}_\eff \equiv \delta\,M_*^3~.
\eeq%
Recalling that the deficit angle in (\ref{xixideficit}) arises
from the extrinsic curvature and that the jump in $R'$ comes from
the intrinsic curvature, we readily identify that
$\sigma^{(4D,5D)}_\eff$ contain a contribution from $K_{\mu\nu}$
and from $G_{\mu\nu}$ respectively, aside from the DW tension
itself. Also, in terms of $\sigma^{(4D,5D)}_\eff$, the wall
junction condition (\ref{xixideficit}) can be expressed as a `sum
rule',
\beq\label{sumrule} %
\sigma^{(4D)}_\eff+\sigma^{(5D)}_\eff=\sigma~.
\eeq %
Note also that the notion of the 5D tension (\ref{5Dsigma}) is not
valid for super-critical walls ($\sigma > 2\pi M_*^3 $), simply
because the deficit angle cannot exceed $2\pi$ without adding
extra sources. The appropriate notion in this case will require
some previous familiarization with super-massive codimension-2
objects, and will be discussed at length in Section
\ref{sec:supermassive}.

\begin{table}[!htb]
\centering%
\begin{tabular}{c|c|c|c}
  branch & DW tension & $\sigma^{(4D)}_\eff/ \sigma$ & $\sigma^{(5D)}_\eff/\sigma$ \\[3mm]
\hline\hline
~&~&\\[-2mm]%
CB & $\sigma < 2\pi M_*^3$ &  $0$  & $1$ \\[3mm]
\cline{2-4} %
~ & $\sigma > 2\pi M_*^3$ &  $\ds 1-{2\pi M_*^3\over
\sigma} $ & $\ds {d\over 2r_c}$ \\[3mm] %
\hline %
SAB & $\sigma \ll 2\pi M_*^3$ &  $\ds 2\lp( 1-\lp(\sigma /
M_*^3\rp)^2/48  +\dots \rp)$ & $\ds
- \lp( 1- \lp({\sigma/ M_*^3}\rp)^2/24 +\dots \rp)$ \\[3mm]%
\cline{2-4} %
~ & $\sigma \gg 2\pi M_*^3$  &  $\ds 1+{2\pi M_*^3\over \sigma}
+\dots $ &$\ds -{d\over 2r_c} $
\end{tabular}
\caption{Summary of the screening properties of our solutions. See
the discussion in the text for the definitions of
$\sigma^{(4D)}_\eff$ and $\sigma^{(5D)}_\eff$. Notice that for
sub-critical walls, the sum rule (\ref{sumrule}) is satisfied.}%
\label{table}
\end{table}

Now, let us briefly go through the results displayed in Table
\ref{table}. As mentioned in the introduction, in the Conventional
Branch, sub-critical walls suffer a complete screening of their 4D
tension. The wall effectively switches on the extrinsic curvature
in such a way that it completely compensates for the wall tension.
Let us emphasize that the 4D curvature of the brane is zero in
this case even on the DW location. Precisely for this reason, the
5D tension is not screened at all.

Still for the Conventional Branch, the supercritical walls show a
very different behaviour. First of all, the wall starts to
inflate. So, the screening of the 4D tension is not total. At the
same time, the 5D tension is screened by a huge factor. What
happens in this case is that the space transverse to the wall is
compactified to a size of the order of the thickness $d$ of the DW
core (we are assuming that $d$ is finite but small, $d\ll r_c$).
Hence, the model changes significantly, {\em e.g.}, there exists a
graviton zero mode.
Nevertheless, as we shall show in Section \ref{sec:supermassive},
`ordinary' super-massive codimension-2 objects ({\em i.e.},
without the DGP brane) also compactify one direction of the
transverse space, and start to inflate. Hence, the 5D effect to
which we should compare is the worldvolume inflation rate, rather
than the deficit angle (which is saturated to $2\pi$). As we shall
see, the walls in a DGP brane inflate at a rate suppressed by
$d/2r_c$ relative to the ones without the brane. As argued before,
the analog of $r_*$ scale for for all kinds of walls coincides
with the thickness $d$. Hence, quite surprisingly, the
super-critical walls display the \emph{same} screening of the 5D
tension/mass found in \cite{gi} for the Schwarzchild-like objects.

In the SA branch, we will see that the deficit angle is negative,
meaning that the 5D tension  is 'over-screened' (the sign of the
net tension $\sigma^{(5D)}$ is reversed). On the other hand, from
the 4D point of view, there is anti-screening: the extrinsic
curvature takes the form of a positive tension wall. For small
tensions, $K_{\mu\nu}$ mimics a wall with almost exactly the same
tension, so the 4D effect is enhanced by a factor~2. In the
super-critical case, we shall find a suppression of the Hubble
rate on the wall $\sim d/2r_c$, while the deficit angle is close
to $-2\pi$ instead of $2\pi$ (which is what the sign in the
lower-right entry of Table \ref{table} stands for).

\subsection{Conventional Branch}
\label{sec:CB}

This branch is defined by requiring that the brane extrinsic curvature be
positive ($\epsilon=1$) for non-negative brane tension $\tau$. As seen from
(\ref{deltabeta}), this implies that a positive tension wall
generates as usual a deficit angle in the bulk. From
(\ref{xixideficit}), we also see that this can be
interpreted from the point of view of the brane as a screening
due to the extrinsic
curvature term, which behaves like a negative tension domain wall.
As we will now see, the amount of screening depends on the brane
tension, and in the extreme case $\tau=0$, there is complete
screening.

\subsubsection*{\textbf{Flat domain wall ($\kappa=0$)}}

\begin{figure}[!htb]
\begin{center}
\includegraphics[width=7cm]{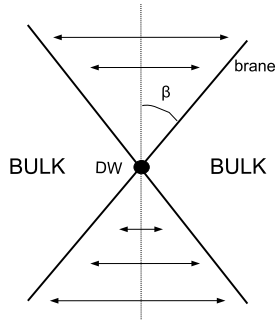} \caption{ Conical
space corresponding to the flat DW case in a flat brane.  The
deficit angle is $\delta=4\beta$, which is always less than
$2\pi$. This figure refers to the Conventional Branch, and with
zero brane tension.}
\label{fig:flatDW}%
\end{center}
\end{figure}

In this case, Eq.  (\ref{angular}) is automatically satisfied if
we set $\tau=0$. On the other hand, Eq. (\ref{xixi1}) is of the
form $R''\propto \sigma \delta(\xi)$. Hence, taking into account
(\ref{Z}), the trajectory is of the form
\beq\label{embedding} %
R(\xi)= |\xi|\;\sin\beta\qquad\qquad Z(\xi)=\xi\;\cos\beta\;
\eeq %
The junction condition on the DW (\ref{xixideficit}) fixes
$\beta$, and leads to an expression for the deficit angle
$$
{\delta}={\sigma\over M^3_*}~,
$$
as for an ordinary codimension-2 object. This space is illustrated
in Fig. \ref{fig:flatDW}. It makes sense as long as
the deficit angle is less than $2\pi$, that is for
$$
\sigma<2\pi M_*^3 ~.
$$
We recognize that the linearized solution of Sec. \ref{sec:lin}
was tailored to reproduce this exact metric.

Now that we have a geometrical method to find the solutions,
we could easily construct multi-center
solutions with as many DWs as we'd like. The induced metric is
always trivial (up to the topology), and the embedding of the
brane in the bulk has a kink-shape, as in Fig \ref{fig:flatDW},
at each DW. In particular, if the wall tensions add up to $2\pi$,
then the bulk is compactified with compactification radius
related to the separation between the walls.

Let us now describe in more detail the critical case, $\sum \sigma
=2 \pi M_*^3$. A single critical wall leads to a seemingly
singular space because the angular component of the metric
(\ref{linsolutionProper}) vanishes. This is of course an artifact
of the thin wall approximation. If we introduce the thickness of
the wall $d$, then the metric asymptotes that of a cylinder of
radius $d$. Hence, the bulk (the transverse space) is
compactified, and in the 'thin wall' limit $d\to0$, one transverse
direction disappears.
 We can illustrate these statements giving
the critical wall some structure, for example by splitting it into
two walls with tensions $3 \pi /2$ and $\pi/2$ (in units of
$M^3_*$), separated by a distance $d$. Since each of them is
subcritical, the bulk is a wedge of $2\pi$ minus the deficit angle
at each wall. In this case, instead of Fig \ref{fig:flatDW}, the
bulk looks like (two copies of) a semi infinite strip of width
$d$.  Thus, the compactification radius coincides with the
separation between walls\footnote{One can  see that for
tensions giving individually deficit angles $\delta_1<\pi$ and
$2\pi-\delta_1$ the compactification radius is $d \sin\delta_1$.}.

This solution contains one modulus, $d$. At first, this is
puzzling because it seems that it costs no energy to move the
walls apart, but this would result in changing the
compactification scale. The reason why there is no paradox is that
to change \emph{globally} the position of the walls actually costs
an infinite amount of energy. The fact that $d$ is a flat
direction means that local perturbations in the wall separation
propagate as massless fields.

There appears to be another paradox, though:  the bulk is
compactified, so one expects to have four dimensional gravity at
distances larger than $d$. However, the DW is flat, so it does not
gravitate as in 4D. This situation occurs also in  ordinary
Kaluza-Klein theories, without the brane. A critical codimension-2
object in 5D compactifies one transverse in the same way,
producing an effectively 4D space. But it does not inflate, as it
would in 4D. This an example of \emph{off-loading}: all the
gravitational effect of the DW tension is 'exhausted' in
compactifying the transverse space. We will see below that if the
tension is supercritical, then it starts behaving as in 4D, and in
the limit $\sigma\gg M_*^3$ we will recover the usual metric of a
domain wall on the brane. But in the critical case, the four
dimensional behavior is not reproduced.

Finally, we shall mention that the space generated by a critical
(distribution of) walls is unstable under nucleation of bubbles of
nothing, very much like with the Kaluza-Klein vacuum
\cite{KKbubble}. The reason is that, the space is asymptotically
cylindrical, which has  the same topology as the bubble of nothing
instanton. The estimated nucleation rate is of order
$\exp(-m_P^2 R^2)$ \cite{KKbubble}, where $R$ is the (asymptotic)
radius of the extra dimension. In our case, this is related to the
wall thickness. For a single critical wall, assuming that its
thickness is of the same order as the scale of the tension $M_*$,
this gives a highly suppressed rate. For an extended distribution of
walls $R$ is larger, so the rate is even more suppressed.

\subsubsection*{\textbf{Inflating domain wall ($\kappa=1$)}}

\begin{figure}[!hbt]
$$
\begin{array}{cc}
  \includegraphics[width=6cm]{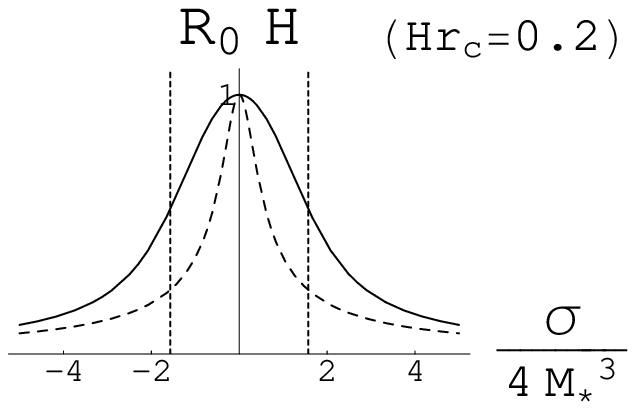}&
  \includegraphics[width=6cm]{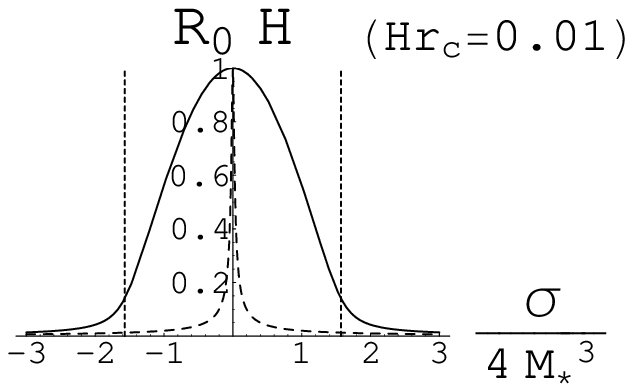}\nonumber\\[.5cm]
  \includegraphics[width=6cm]{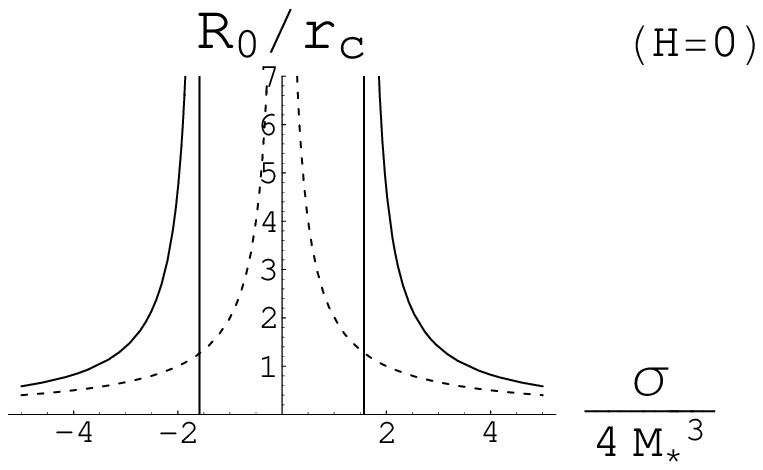}&
    \includegraphics[width=6cm]{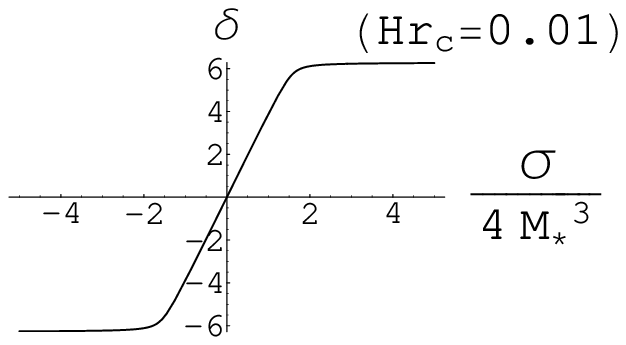}\nonumber\\[-.5cm]
\end{array}
$$
\caption{The radius of the DW $R_0$ for the conventional  branch (CB),
in the $Z_2$ symmetric case, for $H\rc=0.2,0.01$ and $0$, and the deficit
angle (for $H\rc=0.01$) as a function of the tension $\sigma$.
We see how the DW tension is screened by the $K_{\mu\nu}$ term
(contributing effectively negative tension), so that the radius of
the DW is larger than in 4D GR (dashed line).
There is a qualitative change of behavior for  tensions below or
above the critical value $2\pi M_*^3$ (vertical lines). In the
limit $H=0$, tensions smaller than or equal to $2\pi M_*^3$
induce a flat DW.
Only for $\sigma>2\pi M_*^3$, the DW worldsheet is inflating, and
$R_0$ approaches the 4D value ($r_\dw$) for $\sigma\gg M_*^3$.}
\label{fig:Z2Normal} %
\end{figure}

Solving for $\beta$ in Eq. (\ref{xixiBeta}) for $\tau\neq0$, we can
find the deficit angle $\delta$ and the DW radius
$R_0=H^{-1}\cos\beta$. The results are shown in Fig.
\ref{fig:Z2Normal} for $H \rc =1/5$ and  $1/100$. We can see how
as $H\to0$, the subcritical walls (with $\sigma<2\pi M_*^3$) are
much flatter than in usual GR.

For $H=0$, Eq. (\ref{r'r}) implies that $R'^2=1$. Thus, the brane
trajectory lies along $Z=const.$ lines in the $Z-R$ plane, which
means that the deficit angle is $2\pi$ (for the CB). Strictly
speaking, this limit is singular, because the bulk collapses to
zero volume. However, if  we introduced the DW thickness $d$, we
would realize that the bulk is a slice of size roughly given by
$d$. Here, we shall ignore these details and assume that the
geometry is appropriately regularized in such a way that it still
makes sense to speak of a deficit angle $2\pi$. Then, the matching
condition (\ref{xixideficit}) takes the following form,
\beq\label{supermassive} %
{1\over R_0}+\,{\pi\over 4 \rc}={1\over r_\dw}
\eeq%
The solutions of this equation are represented in the lower left
plot of Fig \ref{fig:Z2Normal}. Solutions with a finite Hubble
radius $R_0$ start to exist for supercritical walls
$\sigma>{2\pi}M_*^3$.  Subcritical walls have an infinite $R_0$,
that is they are flat, in agreement with the arguments of the
previous sections.

\subsection{Self-accelerated branch}
\label{sec:SAB}

In this case, the brane extrinsic curvature is negative
($\epsilon=-1$). This implies that a positive tension wall
generates a negative deficit angle in the bulk (see Fig
\ref{fig:Z2SAB}, or Eq. (\ref{deltabeta})). From the point of view
of the brane, this is seen as an anti-screening due to the
extrinsic curvature term.
Using {\em e.g.} (\ref{xixiBeta}) for $\tan\beta\simeq\beta\ll1$,
one obtains $\tan\beta\propto R'_0/R_0$ twice as large as in GR.
This means that in this case $K_{\mu\nu}$ behaves like a positive
tension domain wall with effective tension $\sigma$.

\begin{figure}[!htb]
$$
\begin{array}{ccc}
  \includegraphics[width=5cm]{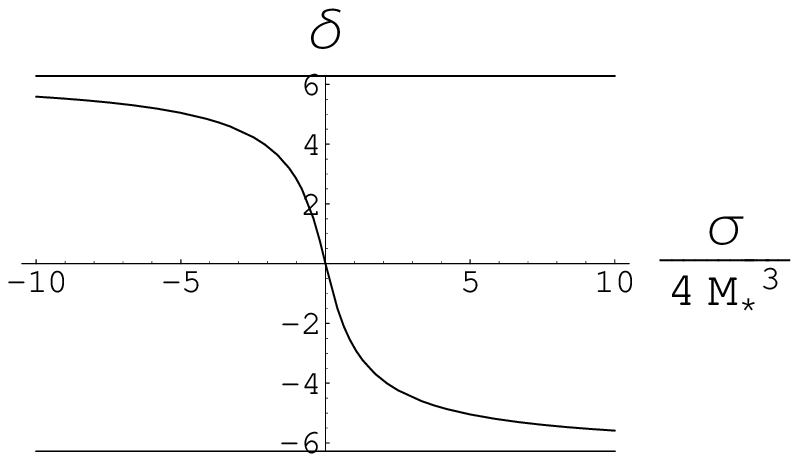} &
  \includegraphics[width=5cm]{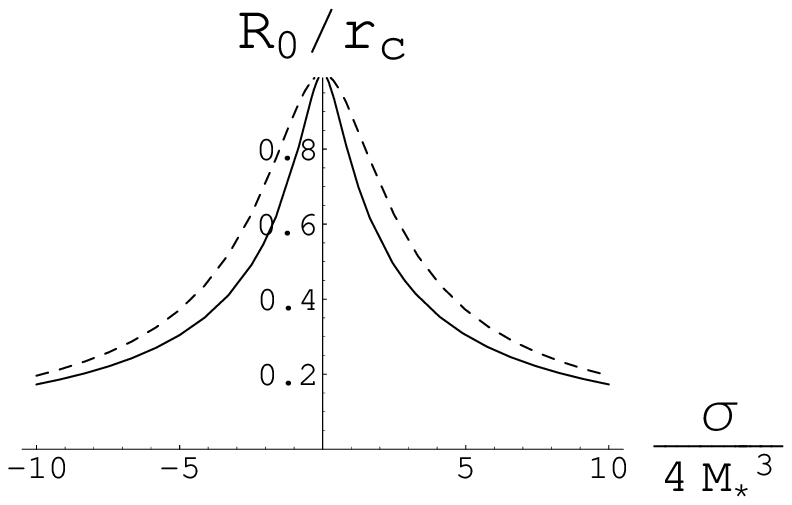} &
    \includegraphics[width=5cm]{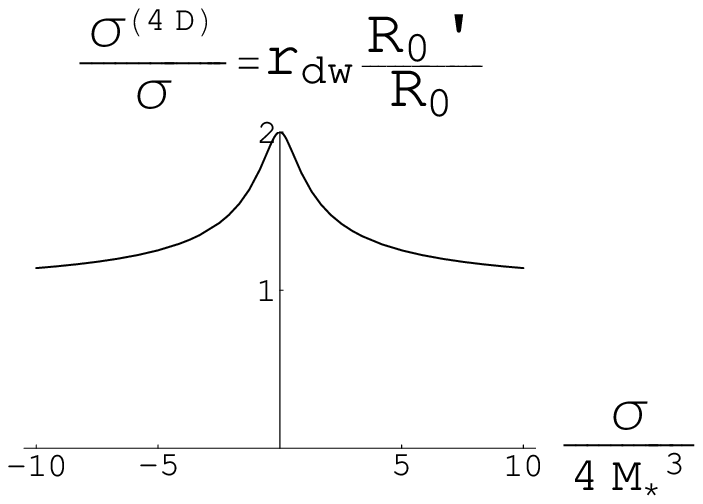}
\nonumber\\[-.5cm]
\end{array}
$$
\caption{The deficit angle and the radius $R_0$  of the DW for the
Self-Accelerated Branch. The radius $R_0$ approaches its 4D value
(dashed line) for large $\sigma$. The fact that $R_0$ is smaller
than the GR result is a consequence of the anti-screening produced
by the extrinsic curvature term. In the third plot, we see the
anti-screened the 4D tension, which amounts to a factor $2$ for
small $\sigma$. For large tension, $\sigma^{(4D)}$ approaches the
GR result.}
\label{fig:Z2self} %
\end{figure}

From the point of view of the bulk, however, there is
`over-screening'. Indeed, the deficit angle is negative (see Fig
\ref{fig:Z2self}), which means that the effective
codimension-2-tension is negative. This can also be visualized
from Fig \ref{fig:Z2SAB}: for the SA branch, the bulk is the
exterior of the brane, which gives an \emph{excess} angle.

We can also see this from Eq (\ref{xixiBeta}) for small DW
tension, $\sigma\ll M_*^3$. In this limit also $\beta\ll1$, so the
deficit angle is approximately (here we take $\kappa=1$)
\beq\label{deficitapprox} %
\delta \simeq {1\over 1+ 2 \epsilon H \rc  } \; {\sigma\over
M_*^3}~.
\eeq%
Hence, in the CB there is screening, while on in the SAB there is
'over-screening'. In fact for the SAB with zero tension, the
deficit angle is exactly opposite as the naively expected.

Two comments are now in order. Given that from the point of view
of the bulk the wall in the SAB behaves as with negative tension,
it appears that the stability of this solution is not granted.
However, a perturbative analysis would not be conclusive, since
the brane is in a strong coupling regime for this solution \cite{dgi,dvali06}.
On the other hand, it is clear that a perturbative treatment as in
Section \ref{sec:lin} will not capture the over-screening. This
is precisely a consequence of the non-perturbative nature of the
solution.

\subsection{Super-critical walls and screening of the 5D tension}
\label{sec:supermassive}

In this section we discuss the spacetime produced by a (local)
supermassive co\-dimension-2 brane in ordinary gravity, in order
to compare it to the supercritical walls found above.
Solutions of a  super-massive  Abelian-Higgs vortex have been
previously found numerically \cite{dLtv,cho}. The metric was shown
to be regular and time-dependent, essentially because
topological-inflation sets in.

In our case, the codimension-2 object is assumed to have the form of
(thick) a Domain Wall
confined on a `fake' brane (with negligible tension and thickness, and no
induced gravity term). As such, it does not have
cylindrical symmetry. Rather, it is like a strip of a codimension-1
object of a finite extent, which, seen from afar appears as
codimension-2. Thus, the problem under consideration is slightly different from
that of \cite{dLtv,cho}. The advantage is that we can find the solution
analytically, and the detailed structure of the core is not going
to be important. Let us also emphasize that the solutions presented
in this section are not the most general compatible with the
assumed symmetries,  since we will also require that
the space outside the object is locally flat
(this is a similar to the restriction to $C_u=0$, $u=1$
in (\ref{Cmetric})).
We shall show that also these ordinary codimension-2
objects with
tension $\sigma>2\pi M_*^3$, the worldsheet  starts to inflate,
and we shall compute at which rate.

We can proceed in the same way as in Section \ref{sec:exact} with
vanishing brane tension $\tau=0$. We shall concentrate on the five
dimensional case, and take a generic form of the energy momentum
tensor compatible with the symmetries of the wall,
$$
T_\mu^\nu={\rm diag}(-\rho,-\rho,-\rho,P)~,
$$
and we assume that $\rho$ and $P$ (if nonzero) depend on $\xi$
only and  are peaked around $\xi=0$. We shall not assume any
particular microscopic model for the wall. Let us just mention
that thick gravitating domain walls arising from scalar field
models do develop a non-vanishing pressure in the core
\cite{widrow} (see also \cite{wang}).

The embedding equations (\ref{angular}) and (\ref{xixi1}) without
the induced gravity term and with  $\tau=0$ become
\beq%
\label{ang}%
2 \kappa\; M_*^3\;{\sqrt{1-R'^2}\over R} =
%-\kappa\;m_P^2\;{1-R'^2\over R^2}\;
-{P\over3}~,
\eeq%
and
\beq%
\label{xx}%
- 2M_*^3\; {R''\over \sqrt{1-{R'}^2}} = %
%2\kappa\;m_P^2\;{R''\over R}+
\lp({2\over3}P+\rho\rp)~,
\eeq%
which in fact is equivalent to (\ref{ang}) once the conservation
equation
$$
P'+3\kappa\,{R'\over R}\lp(P+\rho\rp)=0
$$
is used.

Equation (\ref{ang}) immediately tells us that $P\leq 0$, and that
it should vanish when the worldvolume does not inflate
($\kappa=0$). Instead, if it inflates, then $P$ is nonzero
essentially at the location of the object, where $R'^2\neq1$.

As seen from (\ref{embedding}), a subcritical wall produces a
discontinuity $|\Delta R'|<2$ while a critical wall saturates this
bound. For supercritical walls, then there are two options: one is
to have a larger $|\Delta R'|$, but this would require another
source at finite distance. The other is that the $P$ switches on.
Let us now see that this is compatible with keeping $|\Delta
R'|=2$ for the supercritical walls. The integral of Eq. (\ref{xx})
around the core (that is from $R'=1$ to $R'=-1$) gives
$$
2\pi M_*^3=\int d\xi \,(\rho+2P/3)~.
$$
If we keep calling $\sigma\equiv\int_{core} \rho\,d\xi$ the
tension, and defining the width of the wall $d$ by $\int_{core}
Pd\xi \equiv P_0 \,d$, we obtain
$$
-{2\over3}P_0\,d=\sigma-2\pi M_*^3~,
$$
where $P_0=P(0)$. Now, Eq. (\ref{ang}) tells us that at the center
of the core ($R'=0$) the Hubble rate $1/R_0$ is related to $P_0$
as
$$
-P_0={6 M_*^3\over R_0}
$$
we arrive at
\beq\label{R0supermassive} %
{1\over R_0}={1\over4 d M_*^3}\lp(\sigma-2\pi M_*^3\rp)~.
\eeq %
Given that $R'^2=1$ outside the walls the transverse space is
compactified to a length equal to the thickness $d$. Hence, the
theory becomes effectively 4-dimensional, with an effective 4D Planck mass
$$
M_4^2=M_*^3\,d~.
$$
With this in mind, we can rewrite (\ref{R0supermassive}) as
\beq\label{R0supermassive2} %
{1\over R_0}={\sigma\over4 M_4^2} -{\pi \over 2 d}~.
\eeq %
Hence, we see that we obtained the expected results: Eq
(\ref{R0supermassive}) tells us that for tension larger than $2\pi
M_*^3$, the wall starts to inflate; The Hubble rate approaches
that of a codimension-1 object in GR (in terms of the effective
Planck mass $M_4$). The interpretation of (\ref{R0supermassive2})
in 4D terms is also interesting: the 4D tension of the wall is
screened by the higher dimensional effects by an effective mirror
wall with negative tension and an equivalent horizon of order $d$.

It is also worth describing the structure of the space-time
generated by this object. The space is (the interior of)
a flat `pancake' of a thickness essentially given by the
thickness of the wall, $d$. Hence, the picture is similar to
Fig. \ref{fig:Z2SAB}, but with the `sides' completely flat.
The finite expansion rate on the wall
translates into the fact that the object is placed
at a finite radial distance $R$ (recall that in Fig
\ref{fig:Z2SAB}, the horizontal coordinate is a radial Rindler
coordinate). Accordingly, the presence of a horizon at a finite distance
from the object is seen in the picture because the coordinate
orthogonal to the object `terminates' at the origin $R=0$ (which
is actually the horizon).
\\

Now that we know how supermassive codimension-2 objects behave, we
can compare to what we found for the DGP model. Note that
(\ref{R0supermassive2}) is very similar to (\ref{supermassive})
with the thickness $d$ instead of  $2r_c$ and $M_4$ instead of
$m_P$ (the coefficient in the induced gravity term). One
immediately realizes that the Hubble rates induced on the
objects in DGP is extremely suppressed as compared to ordinary 5D
gravity,
\beq\label{5Dscreening} %
{1/R_0^{(DGP)}\over 1/R_0^{(5D)}}={d\over 2r_c}~.
\eeq %
Hence, the 5D tension is suppressed by a factor $d/r_c$, which as
argued before is the same as $r_*/r_c$, since the
scale where the nonlinearities build up is $r_*=d$ for domain walls.
Note that this is exactly the same suppression that occurred in
the Schwarzchild solution found in \cite{gi},  which also applies
to the supercritical case where the radius of the source is less
than its own $r_*$. Note also that for the domain walls, the 4D
tension is screened, even though not completely. Indeed, without
screening of the 4D tension one would expect $1/ R_0=\sigma/
4m_P^2$ and instead one has (see (\ref{supermassive}))
$$
{1\over R_0}={\sigma\over 4m_P^2}-{\pi\over 4 r_c}~.
$$

At this point, it might seem that the screening of the 5D tension
(\ref{5Dscreening}) for super-critical walls is suspiciously
discontinuous: in the 5D limit, $r_c\to0$, (\ref{5Dscreening}) should go to $1$.
Let us now show that this is precisely what happens,
once we include the thickness of the
wall from the beginning when the wall is on a DGP brane. We only
need to repeat the above steps keeping the induced gravity terms.
The junction equations are
\beq%
\label{ang2}%
2 \epsilon \; M_*^3\;{\sqrt{1-R'^2}\over R} =
-\;m_P^2\;{1-R'^2\over R^2}\;-{P\over3}~,
\eeq%
and
\beq%
\label{xx2}%
- 2\epsilon M_*^3\; {R''\over \sqrt{1-{R'}^2}} = %
2\;m_P^2\;{R''\over R}+\lp({2\over3}P+\rho\rp)~.
\eeq%
On one hand, we can read from (\ref{ang2}) that
\beq%
\label{P0}%
-{P_0\over 3} =\lp( {2\epsilon\,M_*^3 \over R_0}+{m_P^2\over
R_0^2}\rp)~.
\eeq %
Integrating this equation along the core of the wall, we obtain,
for the Conventional Branch,
\beq\label{intxx2CB}%
2\pi\, M_*^3 = {2\over3}d\,P_0+\sigma-4 m_P^2\,{1\over R_0}~.
\eeq%
From this and (\ref{P0}), we arrive at
\beq\label{generalCB} %
4{m_P^2+d\, M_*^3 \over R_0}+2 {m_P^2\,d\over R_0^2}
=\sigma-2\pi\,M_*^3~.
\eeq%
Assuming that $d\ll r_c$, one sees that the for $R_0\gg d$ the
$1/R_0$ term dominates. Hence, we obtain
\beq\label{R0CB1} %
{1\over R_0}={1\over4}\,{\sigma-2\pi\,M_*^3\over m_P^2+d\,M_*^3},
\eeq%
which gives the suppression factor
\beq\label{supp} %
{d\over  d+2 r_c} \simeq {d\over 2 r_c}~.
\eeq %
Note that (\ref{R0CB1}) being consistent with our assumption that
$R_0\gg d$ leads to
\beq\label{consistency} %
d \ll {8\over\pi} { 1 \over {\sigma\over 2\pi M_*^3}-1  }\; r_c~.
\eeq %
This, though, does not seem to be much of a constraint for
microscopic walls unless the tension is extremely large.

For the SAB, (\ref{intxx2CB}) reads
\beq\label{intxx2SAB}%
-4\arccos(R_0^2/r_c^2)\, \; M_*^3 = {2\over3}d\,P_0+\sigma+4
m_P^2\,{1\over R_0}
\eeq%
so the analog of (\ref{generalCB})
\beq\label{generalSAB} %
4m_P^2{1-d/2r_c \over R_0}+2 {m_P^2\,d\over R_0^2}
=\sigma+4M_*^3\,\arccos(R_0^2/r_c^2)~.
\eeq%
Again, assuming that $d\ll r_c$ one obtains for $d\ll R_0\ll r_c$,
\beq\label{R0SAB} %
{1\over R_0}={\sigma+2\pi M_*^3\over 4 (1-d/2r_c) m_P^2}~,
\eeq%
leading to the suppression factor
$$
{\sigma+2\pi M_*^3\over \sigma-2\pi M_*^3} {d\over 2 r_c-d}~,
$$
which, for $\sigma\gg M_*^3 $ coincides with the one for the CB,
(\ref{supp}). Also, the consistency of (\ref{R0SAB}) with the
approximation $d\ll R_0\ll r_c$ leads to a mild condition similar
to (\ref{consistency}).

\section{Domain walls in IR modified gravity}
\label{sec:gia}

Our findings of the previous section may well hold in more general
models of modified gravity. Detailed arguments in favor of this
will be discussed below. However, before we plunge into those
considerations we'd like to emphasize the following: Nonlinear
effects played an important role in  the previous section where we
obtained the results in DGP.  Unfortunately, it is not clear
whether the general models which we will discuss below have any
consistent nonlinear counterparts. For one, massive gravity is
known not to have a consistent  nonlinear completion \cite
{BD,GGruz,CedricRom}, as any of its versions gives rise to certain
ghost-like instabilities.  All of our discussions concerning
general models below ignore these difficulties, and we have  no
way  of assessing at present the significance of this for our
discussions.

The most general ghost-free linearized theory in which gravity is mediated by
a symmetric tensor field $h_{\mu\nu}$, is a generalization of the
Fierz-Pauli \cite {PF} massive theory, and satisfies the following
equation \cite{dvali06},
\begin{equation}
\label{pf} \mathcal{E}^{\alpha\beta}_{\mu\nu} h_{\alpha\beta}\, \,
- \, m^2(\square)\, (h_{\mu\nu} \, - \, \eta_{\mu\nu} h) \,  = \,
-16\pi G_N\; T_{\mu\nu}~,
\end{equation}
where,
\begin{equation}
\label{einstein} \mathcal{E}^{\alpha\beta}_{\mu\nu}
h_{\alpha\beta}\, = \, \square h_{\mu\nu} \, - \, \square
\eta_{\mu\nu} h  \, - \, \partial^{\alpha}\partial_{\mu}
h_{\alpha\nu} \, -\,
\partial^{\alpha} \partial_{\nu} h_{\alpha\mu} \, + \,
\eta_{\mu\nu} \partial^{\alpha}\partial^{\beta}h_{\alpha\beta}\, +
\, \partial_{\mu}\partial_{\nu} h ~,
\end{equation}
$h\equiv h_\mu^\mu$ and $G_N$ stands for the Newton's constant.

For $m^2(\Box)=0$ this reduces to the linearized Einstein's
equations, describing the propagation of two polarizations of
a massless spin-2 graviton. Here, we shall be interested mostly in
the case  $m^2(\Box)=const$, that corresponds to massive gravity;
we will compare the latter to $m^2(\Box)=\sqrt\Box/r_c$, which
corresponds to the DGP model. Our conclusions also
apply to the more general cases where
\beq\label{m2box} %
m^2(\Box)=r_c^{-2(1-\alpha)}\Box^\alpha~, %
\eeq %
with $0\leq\alpha<1$ (the lower bound arising from unitarity
\cite{dvali06}), assuming that these models have sensible
nonlinear completions.

Gravitons satisfying (\ref{pf}) propagate 5
polarizations for nonzero $m^2(\Box)$, which is at the root of the
vDVZ discontinuity \cite {vDVZ}.
The {\it naive} linearized metric produced by a
localized source $T_{\mu\nu}$ is given by
\begin{equation}
\label{source} %
h_{\mu\nu}=-16\pi G_N {1\over
\Box-m^2(\Box)}\lp\{T_{\mu\nu}-{1\over3}\lp(\eta_{\mu\nu}-{1\over
m^2(\Box)} \partial_\mu\partial_\nu\rp) T \rp\}~,
\end{equation}
where $T\equiv T_\mu^\mu$. The factor $1/3$ reflects the existence
of the extra helicity-0 polarization $\chi$, residing in
$h_{\mu\nu}$. The latter could be separated as follows:
\beq\label{h_chi}%
h_{\mu\nu}=\tilde h_{\mu\nu}
+{1\over6}\eta_{\mu\nu}\,\chi+{1\over3}{\partial_\mu\partial_\nu\over
m^2(\Box)} \,\chi ~,
\eeq%
where $\tilde h_{\mu\nu}$ contains the usual 2 polarizations, as
in the massless case, and the helicity-1 states are ignored as they do
not couple to conserved sources in the linearized theory.
Convoluting Eq. (\ref{source}) with a stress-tensor of a
test source $T'_{\mu\nu}$, we obtain the following one-graviton
exchange amplitude%
$$
{\cal A}\;\propto\; G_N \int d^4x \;{T_{\mu\nu}
T'^{\mu\nu}-{1\over3}T T'\over \Box-m^2(\Box)}~.
$$
Let us apply this to an infinite domain wall with the
stress-tensor given in (\ref {dwt}).  Taking a non-relativistic
probe $T'_{\mu\nu}=M\, \delta_\mu^0\delta_\nu^0\,\delta^{(3)}(r)$,
we obtain that the amplitude equals to zero! Actually, it is also
easy to see that it vanishes for any conserved source. First, note
that for the DW, the tensorial structure $T_{\mu\nu}-(1/3)T
\eta_{\mu\nu}\propto \delta_\mu^z \delta_\nu^z$. Hence, the
amplitude is
$$
%{\cal A}\;\propto\;
\int d^4x\; T'_{zz} f(z) =%
-\int d^4x\; \partial^z T'_{zz} \int^z dz' f(z') =%
\int d^4x\; \partial^A T'_{Az} \int^z dz' f(z') =%
0~,
$$
where $f(z)$ is the result of applying $(\Box -
m^2(\Box))^{-1}$ on $\delta(z)$ (certainly a function of $z$
only). In the derivation above, we integrated by parts, used the local conservation of
$T'_{\mu\nu}$, and integrated by parts again (the label $A$ stands
for all coordinates except $z$). Hence, we conclude that the amplitude
vanishes, which is due to the `pure gauge' form
of the metric (\ref{source}) for the DWs.

Thus, in the {\it naive} linearized approximation, domain walls in
massive gravity (and in general theories parameterized by (\ref
{m2box})) do not gravitate. This conclusion, however, is not
warranted until the nonlinear effects are taken into account. This
is because of two interrelated reasons already discussed in
Sections 1 and 2: (1) There is an ambiguity in linearized
solutions; one cannot decide whether to choose a time-dependent
or a static solution, without knowing what is going on with the
source itself. (2) In modified gravity, non-linearities introduce
a scale $r_*$ at which the perturbative expansion breaks down,
and, therefore at $r \lsim r_*$ the above results cannot be
applied. In DGP these issues were addressed by finding exact
nonlinear solutions in Section \ref{sec:exact}.  We do not have the
luxury of a stable nonlinear theory in the case of massive gravity
(and the general $\alpha$-theories (\ref {m2box})). However, this
may only be a temporary technical problem, and if so, it would
make sense to try and estimate what  the analog of the $r_*$
scale for a domain wall in massive gravity is. We will do this below.

First we recall that the $r_*$ scale and strong
coupling originates from the  $1/m^2(\Box)$ term in
(\ref{source}), which at the non-linear level gives rise to
interactions that become singular for $m^2(\Box)\to0$ \cite {ddgv}.
In a schematic way, the leading singular vertex has the form \cite{dvali06}%
\beq\label{trilinear} %
 r_c^{4(1-\alpha)} {\partial^4\over \Box^{2\alpha}} \;\chi \; (\partial\chi)^2~,
\eeq %
where we should keep in mind that the four derivatives do not
necessarily come in  the form of $\Box^2$. For $\alpha=0,1/2$, this was
found in \cite{ddgv} (see also \cite{harv} for $\alpha=0$).
Due to this
singularity, the localized gravitating sources end up having a new
`horizon' at $r=r_*$, where  the perturbative expansion in $G_N$
breaks down. Consider now a non-relativistic spherically symmetric
source $T_{\mu\nu}=M\,\delta_\mu^0\delta_\nu^0\,\delta^{(3)}(r)$,
with a gravitational radius $r_g=2 G_N M$. The scale $r_*$ is then
found as follows:  Assume that far enough from the source, the
linearized solution
\beq\label{far}%
h_{\mu\nu}=-
{\delta_\mu^0\delta_\nu^0-{1\over3}\lp(\eta_{\mu\nu}-
{\partial_\mu\partial_\mu\over
m^2(\Box)}\rp) \over \Box-m^2(\Box)}\, 8\pi\,r_g \,\delta^3(r)
\eeq%
is valid. From this and (\ref{h_chi}), one finds that for $r\ll
r_c$
\beq\label{chi_point}%
\chi\simeq {1\over2}\,{r_g\over r}~.
\eeq%
At the scale $r_*$ the contributions from the nonlinear diagrams
(\ref{trilinear}) catch-up with the leading linear one. This
happens when
$$
\lp( {r_c \over r}\rp)^{4(1-\alpha)} {r_g \over r} \sim1~.
$$
This determines the Vainshtein scale \cite {Arkady}
\beq\label{rv}%
r_V \sim \lp(r_g m^{-4}\rp)^{1/5}\,,  %
\eeq%
for massive gravity, and the $r_*$ scale
\beq\label{r*}%
r_* \sim \lp(r_g r_c^2\rp)^{1/3}\,,  %
\eeq%
for DGP.  At  $r\lsim  r_*$, the metric can be found as
a series in $r/r_c$ \cite {Gruzinov}.

Let us now apply similar considerations to  subcritical domain walls.
Assuming that the linearized solution
\beq\label{sourceDW}%
h_{\mu\nu}=-{\delta_\mu^z\delta_\nu^z+{\partial_\mu\partial_\nu\over
3m^2(\Box)}\over \Box-m^2(\Box)}\;16\pi\,\sigma G_N\,\delta(z)~, %
\eeq%
is valid far-away from the source, for $z\ll r_c$
Eq. (\ref{sourceDW}) leads to
$$
\chi=-8\pi\, G_N \sigma \;|z|.
$$
With such a dependence on $z$  the nonlinearities vanish
due to the structure of the vertices. This is obvious
for the leading singular vertex (\ref{trilinear}),
and is also true for all the other interactions. The key
point is that both for $m^2=const$ and for $m^2(\Box)=\sqrt\Box/r_c$,
all non-linear vertices include at least one $\chi$ (or
$h_{\mu\nu}$) with at least two derivatives.
As a result, for solutions for which $h_{\mu\nu}$ is linear in
$z$, the strongly coupled vertices \emph{vanish} outside of the source.
The only place where they can make  contributions is in the core
of the source itself. Thus, the scale $r_*$, for a subcritical
domain wall,  is determined by the transverse size of the wall itself.
This is reminiscent of a metric of a subcritical Schwarschild source
\cite {gi}, except that the subcriticality condition in the two
cases are quantitatively different.

Finally, let us discuss a similar issue for a  wall with a finite
longitudinal extent. Specifically, consider a disk of radius $L$
located on the plane $z=0$, and let us measure the gravitational
interaction of this `wall' along the $z$ axis. For $r_*\ll z \ll
L$, the solution should be well approximated by (\ref{sourceDW}),
and $\chi\propto|z|$ as before. In this case, we could apply the
same reasoning for strong coupling vertices as we did for an
infinite wall, and could conclude that they vanish outside the
wall as long as $r_*$ is smaller than $L$. Let us get a naive
estimate for  the value of $r_*$ for a finite wall. The total
`mass' of the source is (here we ignore the fact that the domain
wall also has a negative pressure)
$$
M\sim \pi\sigma L^2 \sim \pi\delta M_*^3 L^2~,
$$
where we introduced $\delta\equiv\sigma/M^3_*$, and $M_*$ is the
fundamental Planck scale in the bulk (let us concentrate on the
DGP model for the moment). Then the naive estimate $r_*^{naive}\sim(2 G_N M
r_c^2)^{1/3}$ would suggest
$$
r_*^{naive} \sim \lp(\delta L^2 r_c\rp)^{1/3}~.
$$
If $r_*^{naive} $ is smaller than $L$, then
non-linearities never contribute outside the wall.
This gives the following constraint on the size of the
wall%
\footnote{%
For a general theory of the form (\ref{m2box}) the condition
(\ref{L}) generalizes to $ L\gg \lp({r_c\;\sigma/
m_P^2}\rp)^{1/(3-4\alpha)}\,r_c ~. $ It is interesting
that for $\alpha=3/4$, this does not lead to any constraint at all.}
\beq\label{L}%
L\gg \delta \,r_c ~.%
\eeq%
Hence, in order to avoid the strong coupling of the longitudinal
mode $\chi$, the wall radius has to be quite large; for $\delta\sim 1$
it should be larger than $r_c$. This is consistent
with the expectation that the behavior of a DW here
is different from that in GR. The above follows from the fact that
the infinite
domain wall is already probing the scale of modification of
gravity $r_c$.

\section*{Acknowledgments}

We would like to thank Jose Juan Blanco-Pillado,
Alberto Iglesias, Alex Kovner, and Michele Redi
for useful discussions. We also benefited from several discussions
with Keisuke Izumi, Kazuya Koyama and Takahiro Tanaka.
GD and OP also thank the Perimeter
Institute where part of this work was done, and the
participants of the `IR modifications' workshop, Nima Arkani-Hamed,
Sergei Dubovsky and Alberto Nicolis for their
feedback.
GD is supported in part  by David and Lucile  Packard Foundation
Fellowship for  Science and Engineering, and by NSF grant
PHY-0245068.
The work of GG was supported in part by NASA Grant NNGG05GH34G,
and in part by NSF Grant PHY-0403005. OP acknowledges support from
Departament d'Universitats, Recerca i Societat de la
Informaci{\'o} of the Generalitat de Catalunya, under the Beatriu
de Pin{\' o}s Fellowship 2005 BP-A 10131.

\end{document}